\documentclass[prl,twocolumn,longbibliography,superscriptaddress,amsmath,amssymb]{revtex4-2}

\usepackage{graphicx}
\usepackage{dcolumn}
\usepackage{bm}
\usepackage[usenames]{color}
\usepackage[dvipsnames]{xcolor}
\usepackage{placeins}
\usepackage{enumitem}

\usepackage[normalem]{ulem}

\usepackage[linkcolor=blue,urlcolor=blue,citecolor=blue,colorlinks=true]{hyperref}

\newcommand{\rv}  {{\mathbf r}}
\newcommand{\rV}  {{\mathbf r}_\parallel}

\newcommand{\kn}  {{k_\perp}}
\newcommand{\kV}  {{\mathbf k}_\parallel}
\newcommand{\GV}  {{\mathbf G}_\parallel}
\newcommand{\EM}  {m_\text{eff}}
\newcommand{\LEED}{\Phi_\text{\sc leed}^*}

\newcommand{\PE}  {{\Psi_\text{\sc pe}}}

\newcommand{\them}[1]{{\color{blue} ``#1''}}

\newcommand{\KK} {{\mathbf k}_\parallel}

\begin{document}
\author{E. E. Krasovskii}
\affiliation{Universidad del Pais Vasco/Euskal Herriko Unibertsitatea, 20080 Donostia/San Sebasti\'{a}n, Basque Country, Spain}
\affiliation{Donostia International Physics Center (DIPC), 20018 Donostia/San Sebasti\'{a}n, Basque Country, Spain}
\affiliation{IKERBASQUE, Basque Foundation for Science, 48013 Bilbao, Basque Country, Spain}
\author{R.O. Kuzian}
\affiliation{Donostia International Physics Center (DIPC), 20018 Donostia/San Sebasti\'{a}n, Basque Country, Spain}
\affiliation{Institute for Problems of Materials Science NASU, Krzhizhanovskogo 3, 03180 Kiev, Ukraine}
\title{Comment on ``Distinction of Electron Dispersion in Time-Resolved Photoemission Spectroscopy''}

\begin{abstract}
  In a recent paper [Phys. Rev. Lett. {\bf 125}, 043201 (2020)]
  (Ref.~\onlinecite{Liao2020}) Liao {\it et al}. propose a theory of
  the interferometric photoemission delay based on the concepts of the
  photoelectron phase and photoelectron effective mass.  The present
  comment discusses the applicability and limitations of the proposed
  approach based on an {\it ab initio} analysis supported by vast
  literature. Two central assumptions of the paper are questioned,
  namely that the photoelectron can be characterized by a phase (have
  a well-defined phase velocity), and that it can always be ascribed
  an effective mass. Theories based on these concepts are concluded to
  be inapplicable to real solids, which is illustrated by the example
  of the system addressed in Ref.~\onlinecite{Liao2020}.  That the
  basic assumptions of the theory are never fulfilled in nature
  discredits the underlying idea of the ``time-domain interferometric
  solid-state energy-momentum-dispersion imaging method'' suggested in
  Ref.~\onlinecite{Liao2020}. Apart from providing a necessary caution
  to experimentalists, the present comment also gives an insight into
  the photoelectron wave function and points out problems and pitfalls
  inherent in modeling real crystals.
\end{abstract}
\date{\today}

\maketitle

\subsection{Critical notes on the use of the concepts of the effective mass and
propagating phase in photoemission from solids} 

\begin{figure*}[t]
\includegraphics[width=0.95\textwidth]{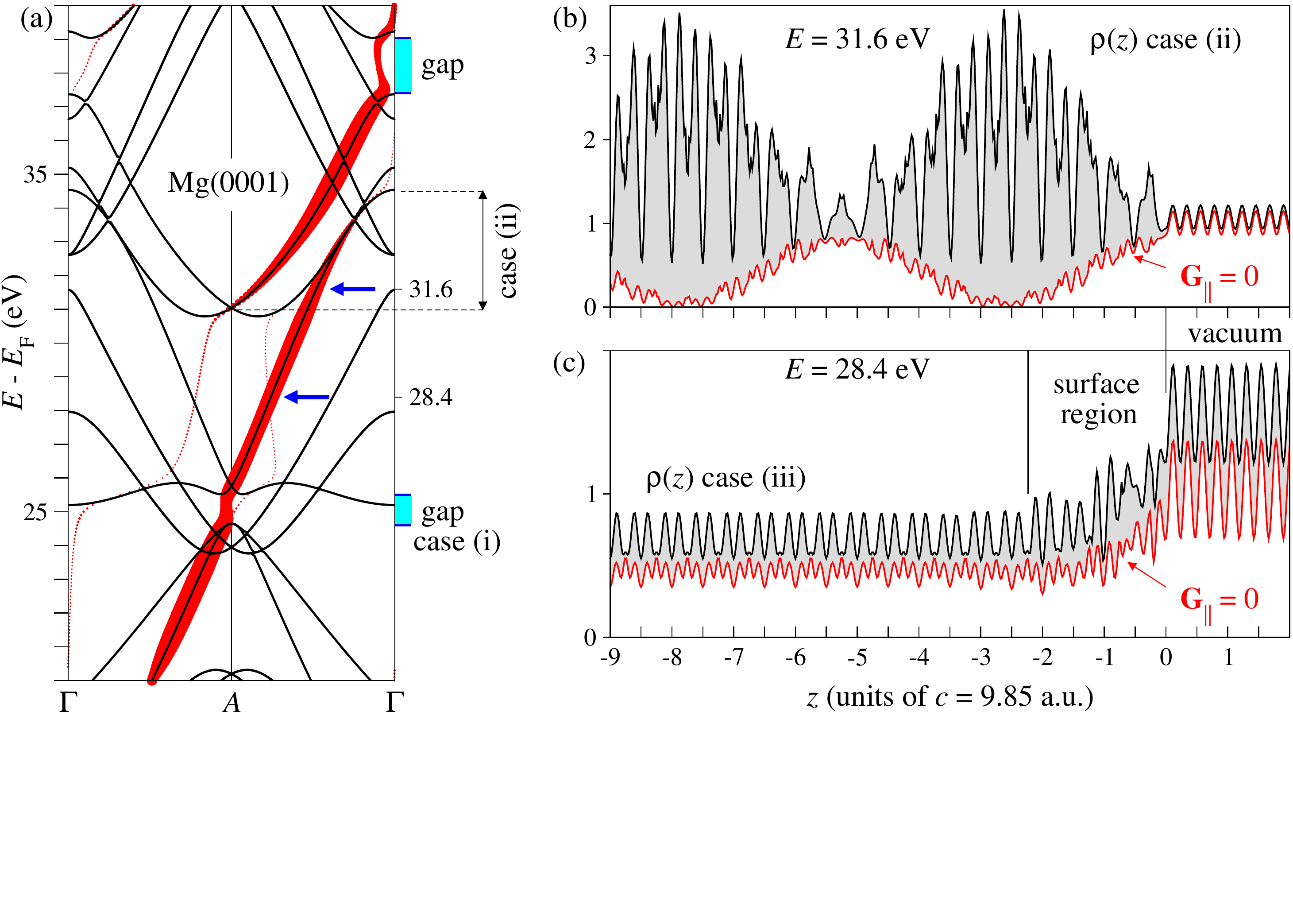}
\caption{(a)~Black lines: real band structure of Mg along $\Gamma A\Gamma$. Red:
conducting CBS given by the dispersion ${\rm Re}\,\kn(E)$ of $\LEED$ constituents.
Line thickness is proportional to the flux carried by the respective Bloch
wave~\cite{Krasovskii2007}. LEED is calculated for absorbing potential
$V_{\rm i}=0.2$~eV~\cite{Krasovskii_1999b}. Arrows show the $2q\pm 1$ final states.
(b), (c) Density profile $\rho(z)$ of $\LEED$ (with $V_{\rm i}=0$) for $2q+1$~(b) and
$2q-1$~(c): full density (black) and single-plane-wave $\GV=0$ contribution (red).
Gray area shows the contribution from the surface harmonics $\GV\ne 0$. 
}
\label{FIG}
\end{figure*}

In Ref.~\cite{Liao2020}, it is proposed to relate the interferometric photoemission delay to
the photoelectron (PE) transport time, based on the PE phase $\xi$ and effective mass $\EM$.
Generally, the concept of effective mass is applicable neither to the PE transport in real
solids nor to the PE interaction with light. While formally the effective mass tensor $\EM$
can be defined for any bulk Bloch state~\cite{Slater1949}, for the PE it is undefined in the
following cases: (i)~When the PE energy lies in a bulk gap, so the outgoing wave $\PE$ contains
no propagating waves. (ii)~When several propagating Bloch states contribute comparably to
$\PE$. (iii)~When the PE mean free path (MFP) is small enough so the weight of the evanescent
waves (complex Bloch vector $\kn$) in the surface region relevant for photoemision is comparable
to that of the propagating wave(s) (real $\kn$). The ubiquity of the $\kV$-projected gaps and
the multi-Bloch-wave character of $\PE$ is well documented for
simple~\cite{Kuzian20,Krasovskii08,Krasovskii10,STROCOV2018},
noble and $d$-metals~\cite{Krasovskii_1999b,Lobo-Checa_2011}, layered
dichalcogenides~\cite{Krasovskii2007,Krasovskii07}, etc. (see illustration in Fig.~\ref{FIG}).

These fundamental and essential features are not allowed for in the basic Eq.~(1) of
Ref.~\cite{Liao2020}, which depends on the ratio ``between the effective energy-dependent PE
masses''. In the cases (i)--(iii) this equation becomes inapplicable because either $\EM$ is
undefined or PE is characterized by more than one $\EM$. This is characteristic of all real
solids, as illustrated for Mg in Fig.~\ref{FIG}(a) by the complex band structure (CBS)
\cite{Heine_1963} responsible for the PE transport. The role of each Bloch wave in
photoemission is characterized by its contribution to the time-reversed LEED state $\LEED$ (low
energy electron diffraction)~\cite{Kuzian20}. We consider the same material 
(see Ref.~\onlinecite{Krasovskii2020} for a detailed study of  photoemission 
from Mg up to photon energies of 320 eV) and 
parameters as in Ref.~\cite{Liao2020}: $\hbar\omega_\text{\sc ir}=1.6$~eV, $2q=50$, and 
$2p$ initial states of 50~eV binding energy. The $2q+1$ final state (31.6~eV) consists of two
propagating waves, while the $2q-1$ state (28.4~eV) comprises only one. However, for 28.4~eV,
down to a depth of at least three atomic layers (15~a.u.) the genuinely evanescent waves
\footnote{In the presence of absorbing potential, all $\kn$ have a nonzero imaginary part;
the waves that have complex wave vectors also with a real potential, $V_{\rm i}=0$, are referred
to as genuinely evanescent waves, see Refs.~\cite{Krasovskii08} and \cite{Heine_1963}.}
generated by the surface~\cite{Appelbaum1972} strongly contribute to $\PE$, as is evident
from the density profile of $\LEED$ in Fig.~\ref{FIG}(c). The depth of 15~a.u. is larger
than the MFP, so the evanescent waves are important both in the PE transport and in optical
transitions forming the sidebands. Thus, even though there is only one real $\kn$, an
effective mass cannot be
ascribed to such wave. Furthermore, the strong admixture of the evanescent waves makes the
notion of a ``time delay $\delta\tau_{a_s}$ accumulated by PEs traveling over one lattice
constant'' inapplicable and Eqs.~(7) and (8) of Ref.~\cite{Liao2020} meaningless.

Let us now imagine a situation when neither of the obstacles (i)--(iii) occurs,
and the outgoing wave in the crystal is a single Bloch wave: 
\begin{equation}\label{eq:f}
\PE(\rv) = \sum\limits_{\GV}\sum\limits_{g}\psi_{\GV g}\exp\left[i(\kn+g)z+i\GV\rV\right].
\end{equation}
Here $g$ are surface-normal and $\GV$ are surface-parallel reciprocal lattice vectors.
The theory of Ref.~\cite{Liao2020} assumes that a propagation phase $\xi = k\Delta z$
can be introduced for the PE wave [Eq.~(2) of Ref.~\cite{Liao2020}]. This assumption is
justified only if one of the plane waves $\{g, \GV\}$ dominates the expansion~(\ref{eq:f}). In
reality, this is almost never the case, see, e.g., the density profiles in Figs.~\ref{FIG}(b)
and \ref{FIG}(c): The weight of $\GV\ne 0$ waves is of the same order as (and often larger
than) the weight of the $\GV=0$ component. Thus, because each of the
$\GV$-waves acquires its own phase $\xi$ the propagation phase is ill-defined for a Bloch wave. 


The conclusion of Ref.~\cite{Liao2020} that ``interferometric spectroscopy
addresses the material-dependent change of PE phase velocities inside the solids''
contradicts the well-documented multi-plane-wave nature of the Bloch states, and the notion
of effective mass conflicts both with the multi-Bloch-wave structure of the PE and with the
surface sensitivity of photoemission. The concepts of PE phase and effective mass, which are
plausible for one-dimensional models, turn out to be inapplicable to real solids, which makes
the method to extract photoemission delays based on Eq.~(1) of Ref.~\cite{Liao2020} unreliable.

\subsection{Addendum: Critique of the Authors' Response}

A shorter version of this note has been published as a Comment~\cite{Krasovskii2021}. For
the benefit of the reader, we quote the main points of the Authors' Response~\cite{Liao2021}
to the Comment.

\smallskip

\begin{itemize}[leftmargin=0mm,label={$^\bullet$}]
\item{\bf Ref.~\onlinecite{Liao2021}:}
\them{... case (i) is irrelevant since the absence of ``propagating waves'' in
    direction normal to the surface prohibits electron emission.}

In fact, electron emission in the absence of propagating waves is not
prohibited. This is known since early days of photoelectron
spectroscopy~\cite{FeibelmanEastman_1974} and is referred to as ``band
gap emission''. It is well-documented experimentally~\cite{COURTHS1979}
and is discussed in classical textbooks~\cite{Feuerbacher1977,Hufner2003}:
{\em ``it is a matter of common experience that EDCs in the UPS regime are also observed when
the final state is in the gap.''}---H\"ufner~2003~\cite{Hufner2003}, p.~383.

\item{\bf Ref.~\onlinecite{Liao2021}:}
\them{With regard to
case (ii), we point out that outgoing final waves are
specified by their momentum (vector), not energy.
Should several propagating Block states with the same
energy contribute, they must have different momenta, both
inside [as shown in Fig. \ref{FIG}(a) of the Comment at 31.6 eV]
and outside the solid, and thus do not contribute to the same
outgoing wave. Such superpositions can be disentangled in
time- and angle-resolved photoemission spectra.}

This statement is 
incorrect. No experiment can distinguish
between photoelectrons excited into different Bloch waves at the same energy and same $\KK$.
This is especially obvious in the one-step theory of photoemission, in which the final state
is represented by the time-reversed LEED
state~\cite{FeibelmanEastman_1974,Freericks2009,Braun2015,Kuzian20}, so the photoemission
matrix element is a coherent sum of the contributions from different Bloch waves, see a
typical example in Fig.~\ref{fig2} below. This gives rise to an interference between
transitions to different Bloch waves, and the respective interference terms have been
thoroughly studied in Ref.~\cite{Krasovskii2007}. The presence of the interference terms
makes it principally impossible to ``disentangle'' the individual contributions in
angle-resolved photoemission spectra.
\begin{figure}[t]
\includegraphics[width=0.48\textwidth]{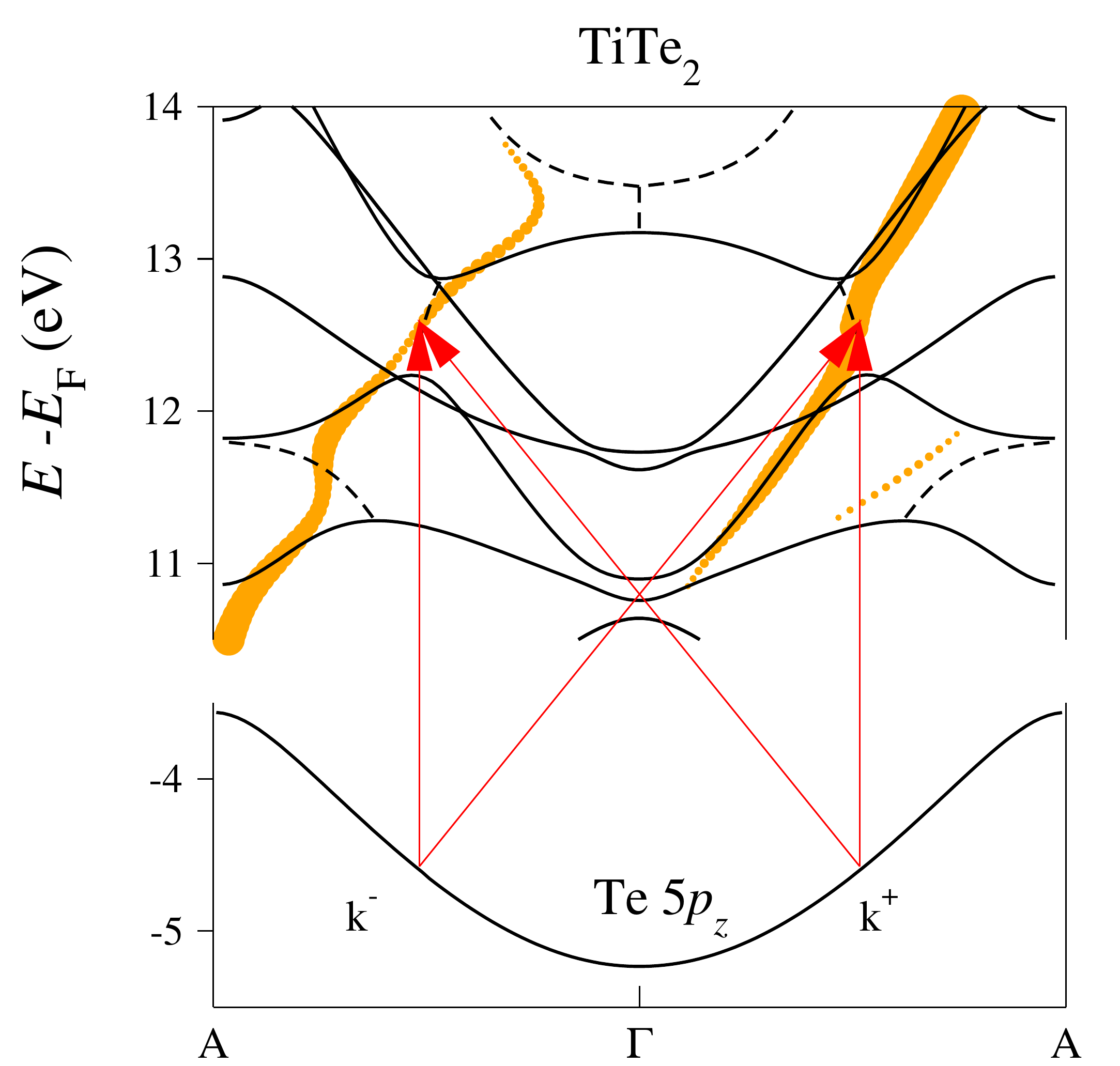}\hspace{2pc}%
\begin{minipage}[b]{0.48\textwidth}\caption{\label{fig2}
    Transitions from the $k^+$ and $k^-$ constituents of the initial Te 5$p_z$ band
    of TiTe$_2$ to Bloch-wave constituents of the unoccupied band structure. Full curves
    display real band structure and dashed curves real lines of complex band structure
    (evanescent waves). Thick lines show conducting constituents of the final states.
    The line thickness is proportional to the current carried (absorbed) by this partial
    wave. The transitions indicated by the red straight lines give rise to interference
    terms in the matrix element, see a detailed discussion in Ref.~\cite{Krasovskii2007}.
    \ \\ \vspace{1mm}
}
\end{minipage}
\end{figure}

Apart from the general criticism above, we have to point at a serious error in the
response by Liao {\it et al}.: They claim that different propagating Bloch states must
have different momenta also {\em outside} the solid. Obviously, the wave that is defined
both inside and outside the solid is not a Bloch wave, and the momentum outside the solid
is uniquely determined by the kinetic energy and the emission direction.

\item{\bf Ref.~\onlinecite{Liao2021}:}
\them{In case (iii), the damped amplitude of the evanescent wave does not directly
influence the phase accumulation during propagation.}

This is not an answer to the case~(iii) remark. Case~(iii) is illustrated
in Fig.~1(c), 
and it concerns the issue of {\em effective mass} rather than
{\em phase accumulation}. Case~(iii) is a generalization of case~(ii), the difference being that
in case~(ii) there are several propagating waves, and in case~(iii) there may be one
propagating and many evanescent waves. For photoemission, propagating and evanescent waves are
equally important, so there is no physical difference between the two cases. However, while
case~(ii) can be excluded at some energies, case~(iii) should be taken into account at every
energy, see the detailed discussions in Refs.~\cite{Appelbaum1972,Bross1982,Wachutka1986}. 

\item{\bf Ref.~\onlinecite{Liao2021}:}
\them{... photoelectron transport in the
solid target while, to our knowledge, photoelectron transport 
times have not been clearly related to the directly
observable phase (difference) information. Equation (1) in
Ref.~\cite{Liao2020} provides this missing link based on the decisively
different effect of photoelectron energy-momentum dispersion 
on the active electron’s phase and group velocity.
}

It is exactly the validity of the concept of {\em wave phase} in
application to photoelectron transport that we question in our Comment, in particular,
the relevance of Eq.~(1) in Ref.~\cite{Liao2020}. The authors made no attempt to respond to our
criticism that the propagation phase is ill-defined for a Bloch wave.

\item{\bf Ref.~\onlinecite{Liao2021}:}
\them{Clearly, the degree to which predictions from one-dimensional models and Eq.~(1)
in Ref.~\cite{Liao2020} are applicable to real (strong-field dressed three-dimensional solids)
remains to be seen.}

It is not clear how the theory of Ref.~\cite{Liao2020} can make predictions
for specific materials, but this is not our point. Whatever remains to be seen, ignoring the
limitations of the model and using the concepts of {\em effective mass} and {\em wave phase}
in cases when they are undefined are essential flaws of the theory.

\item{\bf Ref.~\onlinecite{Liao2021}:}
\them{... one-dimensional numerical models have
been applied with some success to model time-resolved
photoemission from surfaces in the past}

One-dimensional models may be useful and instructive, but they
cannot be used beyond their limitations. 

\item{\bf Ref.~\onlinecite{Liao2021}:}
\them{... we (i) disagree with the conclusion of Krasovskii and Kuzian (KK) that
the concept of photoelectron phase and effective mass ...
are inapplicable to real solids and (ii) find the statement by
KK that for the ``cases (i)---(iii) Eq.~(1) becomes inapplicable'' overstated and inappropriate.}
  
It is important to reveal the limitations and delineate the applicability of the model.
Here, we have questioned the main claim of Ref.~\cite{Liao2020}, namely that
{\em ``attosecond time-resolved interferometric spectroscopy addresses
the material-dependent change of PE phase velocities inside the
solids''}, as well as the main message for experimentalists that
the suggested method can be used {\em ``to extract accurately
photoemission-time delays for imaging electronic dispersion in real solids''}.
\end{itemize}

\subsection{Acknowledgments:}
 This work was supported by the Spanish Ministry of Science and
 Innovation (Project No.~PID2019-105488GB-I00) and by the National
 Academy of Sciences of Ukraine (Project No. III-4-19).

\bibliography{freia}
\end{document}